\documentstyle[12pt]{article}

\advance \textheight by \topskip
\def\ps#1{\raisebox{.2ex}{$\displaystyle
    \mathop{\psi}^{\scriptscriptstyle [#1]}$}{}}
\def\q#1{\raisebox{.2ex}{$\displaystyle
    \mathop{q}^{\scriptscriptstyle (#1)}$}{}}
\def\eps#1{\raisebox{.2ex}{$\displaystyle
    \mathop{\varepsilon}^{\scriptscriptstyle (#1)}$}{}}

\def\th#1{\raisebox{.2ex}{$\displaystyle
    \mathop{\theta}^{\scriptscriptstyle [#1]}$}{}}
\def\Lam#1{\raisebox{.2ex}{$\displaystyle
    \mathop{\Lambda}^{\scriptscriptstyle [#1]}$}{}}

\def\l#1{\raisebox{.2ex}{$\displaystyle
    \mathop{\lambda}^{\scriptscriptstyle [#1]}$}{}}
\def\g#1{\raisebox{.2ex}{$\displaystyle
    \mathop{g}^{\scriptscriptstyle [#1]}$}{}}
\def\Ph#1{\raisebox{.2ex}{$\displaystyle
    \mathop{\Phi}^{\scriptscriptstyle [#1]}$}{}}

\def\A#1#2{{\mathop{A}\limits^{[#1]}_{[#2]}}{}}

\makeatletter
\@addtoreset{equation}{section}

\makeatother

\begin{document}

\begin{titlepage}
\hbox to \hsize{\hfil hep-th/9512167}
\hbox to \hsize{\hfil May, 1994}
\vfill
\large \bf
\begin{center}
GAUGE INVARIANT SYSTEMS OF A GENERAL FORM: \\
COUNT OF THE PHYSICAL DEGREES OF FREEDOM\footnote{Talk presented at the
8th International Seminar {\it Quarks'94}, May 11-18, 1994, Vladimir,
Russia;\\ pp. 486-491 of the Proceedings,
eds.: D. Yu. Grigoriev, V. A. Matveev et al. (World Scientific, 1995)}
\end{center}
\vskip1cm
\normalsize
\begin{center}
{\bf Kh. S. Nirov\footnote{E--mail: nirov@ms2.inr.ac.ru}}\\
{\small \it Institute for Nuclear Research of the Russian Academy of
Sciences,} \\
{\small \it 60th October Anniversary prospect 7a, Moscow 117312, Russia}
\end{center}

\vskip2cm
\begin{abstract}
\noindent
The relationship between various methods to calculate the physical
degrees of freedom for gauge invariant systems of a general form is
established. The set of hidden parameters caused for the superfluous
degrees of freedom is revealed.
\end{abstract}
\vfill
\end{titlepage}

{\bf 1.} In spite of more than 40 years of history of the constrained
dynamics and vast list of publications appeared from the first paper
by P.~Dirac \cite{Dir} there still are lacunas in this theory even on the
classical level. First of all, it concerns the correspondence between
Lagrangian and Hamiltonian formulations of the theory and geometrical
meaning of the Legendre transformation for the degenerate case
(see \cite{Pavlov,Pons,PR} and refs. therein).
Another question deals with the general form of gauge symmetry
transformations and corresponding calculation of physical degrees of
freedom \cite{GrP,HTZ,P}. At the quantum level of the theory we get the
problem of consistent description of the state space of gauge-invariant
systems within the framework of any quantization scheme, e.~g.
BRST-quantization. This problem is still far from completion \cite{NR4}.
In this report we will deal with the count of the physical (dynamical)
degrees of freedom. We suppose the summation over repeated indexes.

\vskip3mm

{\bf 2.} Let us consider the mechanical system given by the Lagrangian
$L(q,\dot q)$, which is invariant under gauge transformations
of the form
\begin{equation}
\delta_\varepsilon q^r = \sum_{k=0}^N
{\eps{k}^\alpha \ps{N-k}^r_\alpha(q,\dot q)},
\qquad r = 1,\ldots,R,
\label{1}
\end{equation}
where $\varepsilon^\alpha(t)$, $\alpha = 1,\ldots,A,$ are infinitesimal
arbitrary functions of time. The velocity phase space $V$ is described
by the set of generalized coordinates $q^r$ and generalized velocities
$\dot q^r$. To calculate correctly the dynamical (physical) degrees of
freedom of the system, it is necessary to consider the gauge
transformations of all coordinates of the velocity phase space point
$\{ \delta_\varepsilon V \}$ =
$\{ \delta_\varepsilon q^r;\delta_\varepsilon \dot q^r \}$ on the
trajectory of the system, which is defined by the Lagrange equations
\begin{equation}
L_r(q,\dot q,\ddot q) \equiv W_{rs}(q,\dot q)\ddot q^s - R_r(q,\dot q)
:= 0,
\label{2}
\end{equation}
where
\begin{eqnarray}
R_r(q,\dot q) = \frac{\partial L(q,\dot q)}{\partial q^r} -
\dot q^s\frac{\partial^2 L(q,\dot q)}{\partial q^s\partial \dot q^r},
\qquad
W_{rs}(q,\dot q) =
\frac{\partial^2 L(q,\dot q)}{\partial \dot q^r \dot q^s},
\label{3}
\end{eqnarray}
and the symbol $:=$ means that the corresponding equation is valid on
the trajectory satisfying the relations $\dot q^r(t) = d/dt q^r(t)$.
Because of the gauge invariance the Hessian of the system $W_{rs}$
is a singular matrix, and we get that the vectors $\ps{0}^r_\alpha$ are
null--vectors of this matrix. Suppose that any null--vector of the
Hessian is a linear combination of $\ps{0}^r_\alpha$. Hence, we can
express the generalized accelerators $\ddot q^r$ from the Lagrange
equations (\ref{2}) only in the following form
\begin{equation}
\ddot q^r := W^{rs} R_s + x^\alpha \ps{0}^r_\alpha , \label{4}
\end{equation}
where $W^{rs}$ is some pseudo--inverse matrix \cite{R}
for the Hessian $W_{rs}$, and $x^\alpha$ are undefined arbitrary parameters.
Further, it is easy to show that the vectors entering the gauge symmetry
transformations satisfy the relations
\begin{equation}
\ps{0}^s_\alpha \frac{\partial \ps{N-k}^r_\beta}{\partial \dot q^s}
= \A{k}{k+1}^\gamma_{\alpha \beta}\,\,\ps{0}^r_\gamma ,
\qquad k = 0,1,\ldots,N ,
\label{5}
\end{equation}
where $\A{k}{k+1}^\gamma_{\alpha \beta}$ are some functions of
the velocity phase space coordinates. Using Eqs.(\ref{1})-(\ref{5})
we get
\begin{equation}
\delta_\varepsilon \dot q^r = \sum_{k=0}^N \eps{k}^\alpha \left(
\ps{N-k+1}^r_\alpha + T(\ps{N-k}^r_\alpha) \right) +
\left( \eps{N+1}^\alpha + x^\beta \sum_{k=0}^N \eps{k}^\gamma
\A{k}{k+1}^\alpha_{\beta \gamma} \right) \ps{0}^r_\alpha , \label{6}
\end{equation}
where $T$ is the differential operator of the form
\begin{equation}
T = \dot q^t \frac{\partial}{\partial q^t} + R_s W^{st}
\frac{\partial}{\partial \dot q^t} . \label{7}
\end{equation}
On the trajectory of the system we have that
\begin{equation}
T := \frac{d}{dt} , \label{8}
\end{equation}
hence, the operator $T$ has the sense of the time evolution operator
of the
 gauge invariant systems.

In order to calculate the physical degrees of freedom we must reveal
all arbitrary parameters of the system and fix them for some initial
value of time $t = t_0$. Upon this, the gauge transformations are
treated as the transformations of the initial data, where
$\eps{k}^\alpha$ form the set of independent arbitrary parameters.

So, let us analyze Eq.(\ref{5}). It can be shown that \cite{N1}
\begin{equation}
\A{k}{k+1}^\gamma_{\alpha \beta} = 0 \qquad {\rm for} \quad k > 1 .
\label{9}
\end{equation}
Hence, from Eq.(\ref{5}) we obtain  the differential equations of the
form
\begin{equation}
\ps{0}^s_\alpha \frac{\partial
\ps{k}^r_\beta}{\partial \dot q^s} = 0 , \qquad k = 0,1,\ldots,N-2 .
\label{10}
\end{equation}
It means that the functions $\ps{k}^r_\alpha(q,\dot q)$, $k =
0,1,\ldots,N-2$, take constant values on the surfaces $S_0$ having
parametric representation of the form
\begin{eqnarray}
q^r(\l{k}) &=& q^r , \label{11} \\
\dot q^r(\l{k}) &=& \dot q^r + \l{k}^\alpha \ps{0}^r_\alpha(q,\dot q) .
\label{12}
\end{eqnarray}
One can show that the arbitrariness in definition of nonzero functions
$\A{0}{1}^\gamma_{\alpha \beta}$,  $\A{1}{2}^\gamma_{\alpha \beta}$
is fixed by fixing of $[x]$-parameters, whereas the arbitrariness of
the vectors $\ps{k}^r_\alpha$, $k = 0,1,\ldots,N-2$, is fixed by
fixing of $[\lambda]$-parameters in Eqs.(\ref{11}),(\ref{12}).

Now taking into account the relations (\ref{1}), (\ref{4}), (\ref{6}),
(\ref{11}), (\ref{12}) we see that to determine the physical degrees
of freedom we must fix $(N+2) \times A$ $[\varepsilon]$-parameters,
$A$ $[x]$--parameters, $(N-1) \times A$ $[\lambda]$--parameters.
Hence, we get the number of the physical degrees of freedom of the
system to be equal
\begin{equation}
2 F_L = 2 R - 2 (N + 1) \times A . \label{13}
\end{equation}

\vskip3mm

{\bf 3.} Another way to obtain the number of the physical degrees
of freedom is the following.
Consider the space of trajectories of the system, where
$\dot q^r(t) = dq^r(t)/dt$. In this scheme $\eps{k}^\alpha$ are not
independent parameters. They satisfy the differential equations
$\eps{k}^\alpha(t) = d^k\varepsilon^\alpha(t)/dt^k$.  It can be shown
that the system under consideration has the following hierarchy of the
Lagrangian constraints
\begin{equation}
\Lam{k}_\alpha (q,\dot q) := 0 , \qquad k=1,\ldots,N, \label{14}
\end{equation}
where the functions $\Lam{k}_\alpha (q,\dot q)$ obey the Noether
identities of the form
\begin{eqnarray}
\Lam{k+1}_\alpha &=& \ps{k}^r_\alpha R_r - \dot q^s
\frac{\partial \Lam{k}_\alpha}{\partial q^s} ,
\label{15} \\
\ps{k}^r_\alpha W_{rs} &=&
- \frac{\partial \Lam{k}_\alpha}{\partial \dot q^s} ,
\label{16}
\end{eqnarray}
where
$k = 0,1,\ldots,N$ and $\Lam{0}_\alpha = \Lam{N+1}_\alpha \equiv 0$.
Now to have the correct number of the physical degrees of freedom we must
choose $A$ functions $\varepsilon^\alpha(t)$ and impose $N \times A$
conditions (\ref{14}) on the trajectories of the system.
Hence, we get the
formula for the physical degrees of freedom
\begin{equation}
F_L = R - (N + 1) \times A , \label{17}
\end{equation}
that gives the result coinciding with (\ref{13}).

\vskip3mm

{\bf 4.} It is desirable to obtain the same result from the Hamiltonian
formulation of the theory. Indeed, it can be shown \cite{N1} that in
the Hamiltonian description of our system there appears a set of $A$
primary constraints $\Ph{0}_\alpha$ and a set of $N \times A$
secondary constraints $\Ph{k}_\alpha$, $k = 1,\ldots,N$, of $N$
stages. These functions $\Ph{0}_\alpha$, $\Ph{k}_\alpha$ are
functionally independent and fulfil the relations
\begin{eqnarray}
\Ph{0}_\alpha (V) &=& 0 , \label{18} \\
\Ph{k}_\alpha (V) &=& \Lam{k}_\alpha , \qquad k = 1,\ldots,N,
\label{19}
\end{eqnarray}
in the terms of the velocity phase space coordinates \cite{N1}.
One can show that the Hamiltonian constraints
of the system form the constraint algebra of the first class \cite{N1}.
The explicit form of the constraint algebra has been found
in ref.\cite{N1}
within the framework of the so-called standard extension procedure
presented first in ref.\cite{PR} and developed in
refs.\cite{NR1,NPR,NR2,NR3}.
It is worth to note that the scheme of standard extension allows us to
intersect the "gauge orbits" -- surfaces $S_0$ -- in such a way that,
in particular, one can put the hidden $[x]$- and $[\lambda]$-parameters,
considered above, to be:
\begin{equation}
\l{k}_\alpha = 0, \qquad x^\alpha = \ddot q^r \chi^\alpha_r (q,\dot q) ,
\label{36}
\end{equation}
where the vectors $\chi^\alpha_r$ are dual to the null--vectors
$\ps{0}^r_\alpha$
\begin{equation}
\ps{0}^r_\alpha \chi_r^\beta = \delta^\beta_\alpha . \label{37}
\end{equation}
{}From the other side such a choice of the hidden parameters
(\ref{36}), (\ref{37}) provides us with the following correspondence
between the differential operators in $V$ and $\Gamma$:
\begin{equation}
\frac{\partial}{\partial q^r} \longleftrightarrow
\frac{\partial}{\partial q^r}; \qquad \frac{\partial}{\partial p_r}
\longleftrightarrow W^{rs} \frac{\partial}{\partial \dot q^s},
\label{38}
\end{equation}
where $W^{rs}(q,\dot q)$ is the pseudo--inverse matrix \cite{R}
for the Hessian $W_{rs}(q,\dot q)$. $W^{rs}$ is uniquely
defined by the relations
\begin{equation}
W^{rs} \chi_s^\alpha = 0 , \qquad W^{rt} W_{ts} = \delta^r_s -
\chi_s^\alpha \ps{0}^r_\alpha . \label{39}
\end{equation}

Thus, we get that the following $(N + 1) \times A$ relations
\begin{equation} \Ph{k}_\alpha \approx 0 , \qquad k = 0,1,\ldots,N
\label{20} \end{equation}
restrict the possible values of the canonical phase space coordinates
$\Gamma$.

Besides, the gauge transformations are mapped to the (canonical) phase
space as follows
\begin{eqnarray}
\delta_\varepsilon q^r &:=& \{\,q^r\,,\,G_\varepsilon\,\} , \label{21} \\
\delta_\varepsilon p_r &:=& \{\,p_r\,,\,G_\varepsilon\,\} , \label{22}
\end{eqnarray}
where $G_\varepsilon$ is the linear combination of the constraints
\begin{equation}
G_\varepsilon = \sum_{k=0}^N \g{k}^\alpha (\varepsilon) \Ph{k}_\alpha .
\label{23}
\end{equation}
The functions $\g{k}^\alpha$ depend on the gauge parameters
$\varepsilon^\alpha$ and their derivatives $\eps{k}^\alpha$ up to
$N$-th order. Hence, fixing the initial value of time $t = t_0$ and
treating again Eqs.(\ref{21}), (\ref{22}) as the transformations of the
initial data of the system we see that it is necessary to fix
$(N + 1) \times A$ parameters $\eps{k}^\alpha$, $k = 0,1,\ldots,N$,
and to impose $(N + 1) \times A$ constraints (\ref{20}) on the phase
space coordinates. This procedure gives the number of the dynamical
degrees of freedom to be equal
\begin{equation}
2 F_H = 2 R - 2 (N + 1) \times A . \label{24}
\end{equation}

The same result can be obtained according to the Dirac scheme \cite{Dir}.
Actually, we have the set of $(N + 1) \times A$ constraints of the first
class. Thus, to remove the degeneracy one should impose
$(N + 1) \times A$ gauge conditions on the canonical variables.
This way also leads to Eq.(\ref{24}) for calculation of the physical
degrees of freedom.

Thus, we have obtained the number of the physical degrees of freedom
for the system under consideration using different approaches and
established the mechanism of annihilation of the superfluous (unphysical)
degrees of freedom.

\vskip3mm

{\bf 5.} Let now the vectors $\ps{N-k}^r_\alpha$ depend on higher order
derivatives of the generalized coordinates up to $M$-th order.
Hence, instead of (\ref{1}) we have  the gauge transformations
\begin{equation}
\delta_\varepsilon q^r = \sum_{k=0}^N
{\eps{k}^\alpha \ps{N-k}^r_\alpha(q,\dot q,\ldots,\q{M})}
\label{25}
\end{equation}
to be the local symmetry transformations of the Lagrangian $L(q,\dot q)$.
It follows from the Noether identities
\begin{equation}
\sum_{k=0}^N {(-1)^k \frac{d^k}{dt^k}
\left(\ps{N-k}^r_\alpha L_r\right)} = 0
\label{26}
\end{equation}
that the vectors $\ps{N-k}^r_\alpha$ are determined up to combinations of
the form
\begin{equation}
\th{N-k}^{rs}_\alpha \, L_s , \qquad
\th{N-k}^{rs}_\alpha = - \th{N-k}^{sr}_\alpha . \label{27}
\end{equation}

Thus, from the Noether identities, gauge
algebras and Jacobi identities, using this arbitrariness,
one can conclude that the dependence of the vectors
$\ps{N-k}^r_\alpha$ on higher order derivatives $\q{M}^r$ is effectively
governed by the number $N$ -- maximal order of time derivatives of the
gauge parameters $\varepsilon^\alpha(t)$. Namely, one can get $M = N$.
Moreover, the vectors $\ps{N}^r_\alpha$ are linear in the variables
$\q{N}^r$, quadratic in $\q{N-1}^r$, and so on, whereas the
null--vectors $\ps{0}^r_\alpha$ depend only on the velocity phase
space coordinates $q^r$, $\dot q^r$. Such a dependence on higher order
derivatives gives rise to the additional superfluous arbitrary
parameters of the type discussed above. These additional parameters,
as it turns out, may lead, in general, to violation of the above
correspondence
between Lagrangian and Hamiltonian approaches to calculation of the
physical degrees of freedom. Detailed discussion of this question (see
e.~g. \cite{P} and refs. therein) is beyond the limits of our report.

\vskip5mm

An analysis of the papers cited here in addition to our talk
allows us to point out one more evidence that the classical
aspects of gauge invariant systems are deeply elaborated,
whereas the problems of quantization are up to now solved only for
the simplest (but actually nontrivial \cite{NR4}) case of abelian
gauge symmetry group.

\vskip5mm

{\bf \large Acknowledgements}

\vskip2mm

I am grateful to Profs. A. V. Razumov and F. V. Tkachov for discussions
and support. This research was supported in part by the International
Science Foundation under grant  MP 9000 .

\end{document}